\documentclass[reprint, twocolumn, showpacs, aps, prd, amssymb, amsmath]{revtex4-1}
\usepackage{graphicx}

\def\e{\begin{equation}}
\def\q{\end{equation}}
\def\m{\begin{eqnarray}}
\def\n{\end{eqnarray}}

\def\be{\begin{equation}}
\def\ee{\end{equation}}

\def\mpl{m_{\mathrm{pl}}}

\def\ben{\begin{enumerate}}
\def\een{\end{enumerate}}
\def\ud{\mathrm{d}}
\def\calN{{\cal N}}

\begin{document}
\title{Comment on Asymptotically Safe Inflation}

\author{S.-H. Henry Tye} 
\affiliation{Laboratory for Elementary Particle Physics, Cornell University, Ithaca, NY 14853, USA}

\author{Jiajun Xu}
\affiliation{Laboratory for Elementary Particle Physics, Cornell University, Ithaca, NY 14853, USA}

\begin{abstract}
We comment on Weinberg's interesting analysis of asymptotically safe inflation \cite{Weinberg:2009wa}. We find that even if the gravity theory exhibits an ultraviolet fixed point, the energy scale during inflation is way too low to drive the theory close to the fixed point value. We choose the specific renormalization group flow away from the fixed point towards the infrared region that reproduces the Newton's constant and today's cosmological constant. We follow this RG flow path to scales below the Planck scale to study the stability of the inflationary scenario. Again, we find that some fine tuning is necessary to get enough efolds of inflation in the asymptotically safe inflationary scenario. 
\end{abstract}


\maketitle

{\bf Introduction: }
Among all the approaches to realize quantum gravity, the asymptotic safety of gravity \cite{Weinberg, Niedermaier:2006wt} is a brilliant approach, in which the unltraviolet (UV) behavior of the theory is controlled by a Non-Gaussian fixed point (NGFP) of the renormalization group (RG) flow, with a finite dimensional critical surface of trajectories attracted to the UV fixed point.  Recently, this scenario  has received increasing attention due to the mounting evidence that such a fixed point exists in the UV, and the critical surface turns out to be 3 dimensional \cite{Codello:2007bd, Codello:2008vh, Benedetti:2009rx, Benedetti:2009gn}. 

A natural area to apply asymptotically safe gravity is the early universe \cite{afcosmos}. Recently, the interesting idea of asymptotically safe inflation was investigated by Weinberg in Ref.\cite{Weinberg:2009wa}. Starting with a general covariant theory of gravitation, Weinberg showed that such a theory allows de-Sitter space as a solution to its classical gravitational field equations. Including time dependence in the Hubble parameter naturally introduces instabilities in the de-Sitter solution that can terminate inflation. Assuming the theory is at the fixed point during inflation, and using the known numerical values for the couplings at the fixed point, Weinberg concludes that, in the absence of some fine-tuning, in some known examples with asymptotic safety, inflation ends prematurely without enough number of efolds achieved. 

In this paper, we re-examine Weinberg's analysis. We find that even if the theory admits a NGFP in the UV, the energy scale generically has to be above the reduced Planck scale $\mpl$ in order for the coupling constants to approach the fixed point. However, inflation proceeds at a much lower energy scale, where the Hubble parameter  $H \sim 10^{-5} \mpl$ and the values of the coupling parameters can be quite different from the fixed point values; this may make a difference in terms of how many efolds of inflation one gets. By solving the RG equations, we determine how the couplings flow away from the fixed point, as the energy is lowered below the Planck scale. In the infrared (IR), we use the known experimental values, namely, the Newton's constant $G_N$ and the cosmological constant $\Lambda$ as well as constraints on higher order terms, to determine the particular RG trajectory in the critical surface that flows away from the UV fixed point towards the IR. 
Along this trajectory at $H/\mpl \sim 10^{-5}$, we find that fine tuning is still needed to achieve enough efolds before instability sets in. Compared with Ref.\cite{Weinberg:2009wa}, the fine tuning we find depends on the ratio $H/\mpl$, but is insensitive to the values of couplings at the fixed point. This is the main point of this comment. 
\\

{\bf Setup: }
We start with a general covariant action of gravitation with higher derivative terms as in Ref.\cite{Weinberg:2009wa}, 
\m
\label{action}
&S&=-\int\ud x^4\sqrt{|g|} \big [ g_0 \mu^4   + g_1 \mu^2 R + g_{2a} R^2 \\\nonumber 
&& + g_{2b} R_{\mu\nu}R^{\mu\nu} + \dots \big] ~,
\n
where $\mu$ is the cutoff scale of the theory. We have extracted powers of $\mu$ explicitly to make the coupling constants $g_i$ dimensionless. The dependence of $g_i$ on $\mu$ is suppressed in the notation. 
In principle, matter fields and all higher derivative terms are there from an effective field theory point of view. 
The running of the couplings $g_i$ satisfy the RG equations of the form
\begin{equation}
\mu \frac{\ud g_i}{\ud \mu} = \beta_i (g_j)
\end{equation}
In order for $g_i(\mu)$ to approach a fixed point $g_i^*$ as $\mu \to \infty$, it is necessary that $\beta_i(g_j^*)=0$. \\


{\bf Asymptotically Safe Inflation: }
With rotational and translational symmetries, we start with the flat space FRW metric
\[
\ud s^2 = -\ud t^2 + a^2(t) \ud x_i^2 ~.
\]
The classical gravitational field equation can be solved through the single equation
\begin{equation}
{\cal N} (\mu, t) = 0 ~.
\end{equation}
\begin{eqnarray}
{\cal N}(\mu, t) &\equiv&  -g_0 - 6\mu^{-2} g_1 H^2 \nonumber \\
&& - \mu^{-4} g_{2a} (216 H^2 \dot{H} - 36 \dot{H}^2 + 72 H \ddot{H}) \label{nt} \quad \\
&& + \mu^{-4} g_{2b} (72 H^2 \dot{H} - 12 \dot{H}^2 + 24 H \ddot{H}) + . . . \nonumber
\end{eqnarray}

To search for de-Sitter solutions, we specify the scale factor
\begin{equation}
a(t) = \exp \left( \bar{H} t \right)
\end{equation}
with $\bar{H}$ the constant Hubble parameter. We see that
\begin{eqnarray}\label{hubble}
\frac{\bar{H}}{\mu} = \sqrt{-\frac{g_0}{6g_1}} 
\end{eqnarray}
is a solution of the classical field equation $\calN (\mu, t) = 0$. The Hubble scale is determined by the value of the coupling constants $g_0$ and $g_1$. 

We now discuss how inflation can end in the asymptotically safe scenario. So far we have only considered the Hubble parameter to be a time independent constant. Including the time dependence, we can write
\begin{equation}
H(t) = \bar{H} + \delta H(t) ~.
\label{destab}
\end{equation}
If there exists an unstable mode, then $\delta H(t)$ will grow, thus ending inflation.

With a time dependent $H(t)$, $\calN(\mu, t) = 0$ now requires that
\begin{eqnarray}
c_0(\mu, \bar{H}) \frac{\delta H}{\bar H} + c_1(\mu, \bar{H}) \frac{\delta \dot{H}}{\bar{H}^2} + c_2(\mu, \bar{H}) \frac{\delta \ddot{H}}{\bar{H}^3} + \dots = 0 ~. \quad\quad
\end{eqnarray}
with $c_i$'s Taylor expansion coefficients of $\calN (\mu, t)$ around $\bar{H}$, whose forms are worked out in Ref.\cite{Weinberg:2009wa}. 

If we schematically write
\[
\delta H(t) \sim \exp \left( \xi \bar{H} t \right) ~,
\]
we get
\begin{eqnarray}\label{eq_xi}
c_0 + c_1 \xi + c_2 \xi^2 + \dots = 0 ~. 
\end{eqnarray}
Roots with $Re(\xi) > 0$ will represent instabilities of the solution, and de-Sitter expansion will only last for $1/Re(\xi)$ efolds. If all the roots have $Re(\xi) < 0$, then de-Sitter expansion will be an attractor solution and, ignoring the effect of the inflaton potential, inflation will last forever. \\


{\bf A Specific Example: }
Consider the action
\begin{equation}
 - \int \ud^4 x \sqrt{|g|} \left[ \frac{\mu^2}{g_N} \left(2\lambda \mu^2 - {\cal R} \right)  + \frac{1}{2s} {\cal C}^2 - \frac{\omega}{3s} {\cal R}^2 \right] \label{action1}
\end{equation}
Here $\mu$ is the energy scale, $g_N$, $\lambda$, $\omega$, $s$ are dimensionless parameters. ${\cal R}$ is the Ricci scalar, and ${\cal C}$ is the Weyl tensor. We require $s > 0$ so that the Euclidean functional integral is damping. 


The $\beta$ functions for $s$ and $\omega$ are given by one-loop perturbation theory \cite{Niedermaier:2006wt},
\begin{eqnarray}
\mu \frac{\ud s}{\ud \mu} &=& -\frac{1}{(4\pi)^2}\frac{133}{10} s^2 ~, \label{rg_s} \\
\mu \frac{\ud \omega}{\ud \mu} &=& -\frac{s}{(4\pi)^2}
\left( \frac{5}{12} + \frac{183}{10}\omega + \frac{10}{3}\omega^2 \right) \label{rg_omega} ~, \\\nonumber
\end{eqnarray}
from which we notice that $\omega$ has a stable fixed point at 
\begin{equation}\label{omega}
\omega^* = -0.0228
\end{equation}
and $s$ is asymptotically free.


To simplify the problem, we will fix $\omega = \omega^*$, $s = 0$ and consider the two dimensional flow of $g_N$ and $\lambda$ first. 
The flow of the Newton and cosmological constants have been studied in literature using a functional renormalization group (FRGE) approach. (see Ref.\cite{Niedermaier:2006wt} for a review). 
We can write schematically \cite{Reuter:2007rv}
\begin{eqnarray} \label{gnflow}
\mu \frac{\ud{g_N}}{\ud \mu} &=& 2 g_N - \gamma_1  g_N^2  + {\cal O}(g_N^3) ~, \\
\mu \frac{\ud{\lambda}}{\ud \mu} &=& -2 \lambda + a_1 g_N + a_2 g_N \lambda + a_3 g_N^2 + {\cal O}(g_N^2 \lambda) 
\quad 
\end{eqnarray}
where $a_1$, $a_2$, $a_3$ are known coefficients at the fixed point, which are functions of $\omega$ and $s$. Setting $\omega=\omega^*$ and $s=0$, and using the analysis in Ref.\cite{Niedermaier:2009zz}, we have
\begin{equation}
\gamma_1 = -a_2 = \frac{2u_2^*}{(4\pi)^2} ~, \quad a_1 = \frac{2u_1^*}{(4\pi)^2} ~.  
\end{equation}
with $u_1^* = 1.38$, $u_2^*=0.73$. 
One should note that the relation $\gamma_1 = - a_2$ may not hold if one introduces matter fields into the theory. 

The RG flow can now be easily solved
\begin{eqnarray}
\lambda &=& \frac{(\mu/\mu_\lambda)^{-4}  + u_1^*}{2 (\mu/\mu_0)^{-2} + 2u_2^*} ~, \label{lambda_mu} \\
g_N &=& (4\pi)^2 \frac{(\mu/\mu_0)^2}{1+ u_2^* (\mu/\mu_0)^2} ~. \label{g_mu}
\end{eqnarray}
with $\mu_0$ and $\mu_\lambda$ free parameters. 

So, in the UV limit $\mu \to \infty$, $g_N$ and $\lambda$ flow to the fixed point 
\begin{equation}
\frac{g^*_N}{(4\pi)^2}  = \frac{1}{u_2^*} = 1.37  ~, 
\quad \lambda^* = \frac{u_1^*}{2u_2^*} = 0.95 
\end{equation}

In the IR limit $\mu \to 0$, the Einstein-Hilbert term in the action (\ref{action1}) reads
\[
\frac{\mu^2}{g_N} \left(2\lambda \mu^2 - {\cal R} \right) = \frac{\mu_\lambda^4 - \mu_0^2 {\cal R}}{(4\pi)^2}
 \equiv  \mpl^2 (\Lambda - \frac{ {\cal R}}{2})
\]
where we have demanded that
\begin{equation}
\mu_0^2 = 8 \pi^2 \mpl^2 = \frac{\pi}{G_N} ~, \quad \Lambda = \frac{\mu_\lambda^4}{(4\pi)^2 \mpl^2}
\end{equation}

The cosmological constant $\Lambda$ contributes energy density $\Lambda \mpl^2 \sim 10^{-120} \mpl^4$, which means for positive and small $\Lambda$, the scale $\mu_\lambda$ and $\mu_0$ has to obey
\begin{eqnarray}
\frac{\mu_\lambda}{\mu_0} \sim 10^{-30} ~. 
\end{eqnarray}

{\bf The Hubble Scale during Inflation:}
Using Eq.(\ref{hubble}), during inflation, we have
$H = \sqrt{\lambda/3} \, \mu $. 
At the fixed point $\lambda^* \sim 1$, $H \sim \mu$. However, from observational data, we are pretty confident that 
\[
\frac{H}{\mpl} \ll 1 \quad \Rightarrow \quad \mu \ll \mu_0 ~.
\]
Imposing $H/\mpl \ll 1$ forces the theory to flow away from the UV fixed point. 

Since the theory cannot be at the fixed point during inflation, we want to know how the coupling constants depend on the ratio $H/\mpl$. Away from fixed point, on scales $H_0 \ll \mu \lesssim \mpl$ ($H_0$ being the Hubble constant today), Eq.(\ref{lambda_mu}) gives 
\begin{eqnarray}
\lambda \approx \frac{u_1^*}{2} \left( \frac{\mu}{\mu_0} \right)^2 
\end{eqnarray}
so we have
\begin{eqnarray}
H &\approx& \sqrt{\frac{u_1^*}{6}} \, \frac{\mu^2}{\mu_0} ~, \label{Hmu} \\
\frac{H}{\mpl} &\approx& \pi \sqrt{\frac{4u_1^*}{3}} \frac{\mu^2}{\mu_0^2} \;\sim\; \frac{\mu^2}{\mu_0^2} 
\end{eqnarray}
Using Eq.(\ref{lambda_mu}, \ref{g_mu}), the coupling constants $g_N$ and $\lambda$ are 
\begin{equation}\label{gl_inf}
\lambda \approx \frac{u_1^*}{2}\frac{H}{\mpl} ~, \quad g_N \approx (4\pi)^2 \frac{H}{\mpl} ~. 
\end{equation}
In terms of the ratio $H/\mu$, we find that
\begin{eqnarray}\label{h/mu}
\frac{H}{\mu} \sim \sqrt{\lambda} \sim \sqrt{\frac{H}{\mpl}} \ll 1 ~,  
\end{eqnarray}
so we are in a regime that higher derivatives terms beyond those included in the action (\ref{action1}) are negligible. \\

{\bf Instability of de-Sitter Solution:}
In the absence of the four or higher order derivative terms in the action (\ref{action1}), the de-Sitter solution is stable. The presence of appropriate matter fields will introduce a positive $u_2^*$ into Eq.(\ref{gnflow}) so that $g_N$ has an UV fixed point. Introducing four and higher order derivative terms into the action (\ref{action1}) introduces instability to the de-Sitter solution. A perturbation to the de-Sitter solution will grow, thus destabilizing the inflationary phase. 

To find the unstable mode to $H$, Eq.(\ref{eq_xi}) reduces to
\begin{eqnarray}\label{eq_xi2}
\xi^2 + 3\xi - A = 0 ~, \quad A = -\frac{3s}{2\omega\lambda g_N}
\end{eqnarray}  
If $A \ll 1$, the two roots are $\xi \approx -3$ and $\xi \approx A/3$. The negative $\xi$ does not lead to any instability. If we pick $\xi = A/3$, the de-Sitter phase can last for $3/A$ efolds. For inflation, we need $A \sim 1/20$. 

Using Eq.(\ref{omega}) and Eq.(\ref{gl_inf}) for $\lambda$, $g_N$ and $\omega$, we get
\begin{eqnarray}
A \sim s \left(\frac{H}{\mpl} \right)^{-2} ~. 
\end{eqnarray}
In order to get $A \sim 1/20$, we need, at the inflation energy scale,
\[
s(\mu = H) \lesssim 10^{-1} \left(\frac{H}{\mpl} \right)^{2}
\]
For $H \sim \mu_0 \sim \mpl$, $s$ can be of order unity. 
However, for the realistic inflationary scenario, we require $H/\mpl \sim 10^{-5}$; so $s \lesssim 10^{-11}$, which is close to the bound obtained by Weinberg \cite{Weinberg:2009wa}. 

The parameter $s$ is asymptotically free, 
\begin{equation}
\label{s_af}
s = {s_0}{\big /}\left[{1 + s_0 \frac{133}{160 \pi^2} \ln (\mu/\mu_s)}\right]
\end{equation}
The ${\cal C}^2$ term in the action (\ref{action1}) introduces a massive tensor mode while the ${\cal R}^2$ term introduces a massive scalar mode.  They will modify the gravitational force which is well checked up to distance as small as sub-milimeters. This implies that the extra modes must be massive enough ($m > 10^{-3}$ eV) so these extra forces are Yukawa damped, which goes like $e^{-mr}/r$. Since $m^2 \sim s \mu^2/g_N $, this implies that 
\begin{equation}\label{s_bound}
s > 10^{-60}
\end{equation}
 where coincidentally, $m \gtrsim u_{\lambda}$.
So $s \lesssim 10^{-11}$ is consistent with the bound and may not be hard to arrange in the early universe. It is not clear 
whether the exponentially small $s$ is a fine tuning or not, since other than the bound (\ref{s_bound}), there is no guiding principle for a natural value of $s$.




For $s \sim 10^{-11}$ during inflation, the parameter $s$ will remain small for the whole energy range from $\mu_{\lambda}$ to $\mpl$, due to its asymptotically free property. This justifies our approximation to set $s\to 0$ in the flow equation. At the same time, a tiny $s$ parameter will ensure that the running of $\omega$ is also small according to Eq.(\ref{rg_omega}), which justifies setting $\omega$ at its fixed point value $\omega^*$ in our analysis. \\

{\bf The Issue with the Ghost Pole: }
Because of the fourth derivative terms in the action (\ref{action1}), the graviton propagator contains, in addition to the usual massless graviton, a massive spin two particle of negative residue, i.e. a ghost. The presence of the ghost leads to violation of unitarity and makes the theory inconsistent at quantum level. 

Let us first look at the scale of the ghost pole. Schematically, the ghost propagator takes the form
\[
\frac{-1}{p^2 - m_2^2} ~, \quad m_2^2 = \frac{s}{g_N}\mu^2 ~. 
\]
We now compare the ghost mass $m_2$ with the Hubble scale. Since away from the fixed point, $H/\mu \sim \sqrt{\lambda}$, we have
\begin{equation}
\frac{m_2}{H} \sim \sqrt{\frac{s}{g_N \lambda}} ~.
\end{equation}
Note that the total number of efolds is
\begin{equation}
N_e = \frac{3}{A} = \frac{-2\omega^* \lambda g_N}{s} ~,
\end{equation}
we therefore have
\begin{equation}
\frac{m_2}{H} = \sqrt{\frac{-2\omega^*}{N_e}} \lesssim \frac{1}{10}
\end{equation}
We see that during inflation, the energy scale is above the ghost mass, if the ghost is present at its naive value.

However, the presence of the ghost may be an artifact of the truncation of the theory. If the theory is not truncated, the propagator denominator could have non-trivial forms which removes the ghost pole. The idea is to absorb quadratic 
counterterms (those which contribute to the propagator) by field redefinitions and include all (infinitely many) counterterms generated in the bare action \cite{Gomis:1995jp}. Since the counterterms are all powers of $H/\mu$, the classical solution we find above should still hold as long as $H/\mu \ll 1$, which is imposed by observation. 

Even if the theory is truncated, with running couplings, it is not clear whether the tree level ghost pole will be hit or not \cite{Benedetti:2009rx}. \\

{\bf Matter Fields:} The ghost problem is totally absent if the 4- and higher-derivative terms are not introduced. In the Einstein gravity theory, $g_N$ has a UV fixed point if there is an appropriate set of matter fields so that $\gamma_1 \propto  4n_V + 2n_D -n_S >0$ in Eq.(\ref{gnflow}), where $n_V$ ($n_D$, $n_S$) is the number of vector (Dirac, scalar) fields \cite{Percacci:2005wu}. In this simple example, the instability discussed above is also absent so the ending of inflation in this scenario follows from the properties of the inflaton potential. This is the standard scenario.
\\


{\bf Power Law Inflation:}
Instead of searching for de-Sitter solution, one can also consider power-law expansion with $a(t) \sim t^p \; (p > 1)$ \cite{Bonanno:2010bt}. For a power-law background, we expect that $p \gtrsim 100$, so that the tensor mode $r = 16/p$ satisfies the observational bound. 

We use the action (\ref{action1}) for illustration. To analyze the instability of the power-law solution, we need to expand around $\bar{H} = p/ t$. The only difference is that $\bar{H} = p/t$ is time dependent, so $\dot{H}$ and $\ddot{H}$ cannot be ignored. The resulting instability equation (\ref{eq_xi}) has a positive root (leading order in the large $p$ limit)
\begin{equation}
\xi = - \sqrt{0.51 \frac{H}{\mpl}} + \sqrt{\frac{1.37 s}{H/\mpl} + 0.51 \frac{H}{\mpl}}
\end{equation}
In reaching the above result, we have used Eq.(\ref{gl_inf}) for the couplings away from the fixed points.

If we take $H/\mpl \sim 10^{-5}$, we have
\begin{equation}
\xi \sim -0.002 + 0.002 \sqrt{1 + 2.7 \times 10^9 s} ~.
\end{equation}
We see that if during inflation $s \sim 1$, $N_e \ll 1$. To get enough e-folds, we need at least $s(\mu = H) \lesssim 10^{-7}$. This requirement of an exponentially small $s$ during inflation is again tied to the ratio $H/\mpl \ll 1$, and is insensitive to the values of $g_N$ and $\lambda$ at the fixed points. \\

{\bf Acknowledgements:}
We thank Eanna Flanagan and Gary Shiu for discussion. This work is supported by the National Science Foundation under grant PHY-0355005.

\end{document}